\documentclass[a4paper,11pt]{article}
\pdfoutput=1 

\usepackage{jcappub} 
\usepackage{bigints}
\usepackage[T1]{fontenc} 

\newcommand{\fracc}[2]{\frac{\textstyle{#1}}{\textstyle{#2}}}
\usepackage{color}

\title{\boldmath Feasibility of singularity avoidance for a collapsing object due to a scalar field
}

\author[a,1]{Eduardo Bittencourt,\note{Corresponding author.}}
\author[a,b]{Alan G. Cesar,}
\author[c,d]{Jonas P. Pereira}

\affiliation[a]{Federal University of Itajub\' a, BPS Av. 1303, Itajub\'a, MG, Brazil}
\affiliation[b]{Brazilian Center for Physical Research, Xavier Sigaud St. 150, Rio de Janeiro, RJ, Brazil}
\affiliation[c]{Federal University of Esp\'irito Santo, Fernando Ferrari Av. 514, Vitória, ES, Brazil}
\affiliation[d]{Nicolaus Copernicus Astronomical Center, Polish Academy of Sciences, Bartycka 18, Warsaw, Poland}

\emailAdd{bittencourt@unifei.edu.br}
\emailAdd{alangcesar@gmail.com}
\emailAdd{jpereira@camk.edu.pl}

\abstract{We study the problem of the gravitational collapse of an object as seen by an external observer. We assume that the resultant spacetime is a match of an external Vaidya spacetime with an interior Friedmann-Lema\^itre-Robertson-Walker (FRLW) spacetime of any spatial curvature and with a scalar field both minimally and non-minimally coupled to the metric. With the goal of studying a contracting (collapsing) object, for the initial moment of observation we take that its energy density and pressure are positive, that there are no trapping surfaces, and that the null energy condition (NEC) and the strong energy condition (SEC) are fulfilled. We show that there are many cases where singularities could be avoided for both the minimal and non-minimal couplings, although the contexts for so are very different in both cases. For the minimal coupling, the avoidance of singularities could happen either through evaporation or altogether, triggered by a violation of the SEC for a period of time. For the non-minimal coupling, the complete singularity avoidance happens only if evaporation takes place, and a temporary violation of the SEC does not thwart the formation of singularities. The above results show the relevance of the global (the whole spacetime) validity of energy conditions for the singularity theorems to be applicable; otherwise, the fate of a collapsing star is not known a priori. At the same time, the surface behavior of a collapsing body offers partial diagnostics of what happens in the inaccessible regions of spacetime to external observers. Our analyses suggest that a bounce behavior of the surface of the initially collapsing object is a fingerprint of the SEC violation in its interior, and that could be due to the existence of scalar fields there. }

\begin{document}
\maketitle
\flushbottom

\section{Introduction}

Matching spacetimes is in general a complex task due to the nonlinearity of general relativity. When working with generic matches, surface degrees of freedom are present on a hypersurface (thin shell) separating two spacetimes \citep{poisson_2004}, resulting in numerous possibilities. One concerns neutron stars with phase transitions \citep{PhysRevD.90.123011,2015ApJ...801...19P,2018ApJ...860...12P,2019ApJ...871...47P,2023Univ....9..305P}. Another intriguing possibility is using thin shells to mimic black holes, known as gravastars \citep{2004CQGra..21.1135V,2007CQGra..24.4191C,2015PhRvD..92l4030P}. With the advent of gravitational-wave astronomy, it may be possible to probe them with gravitational waves \citep{2009PhRvD..80l4047P,2010PhRvD..81h4011P}. Thin shells can also be used to construct wormholes \citep{1995PhRvD..52.7318P}. In contrast, when surface degrees of freedom are absent, the matching process becomes more restrictive as it must meet additional constraints. However, this scenario is simpler to describe physically since there is no need to account for fluids (exotic or otherwise) on phase-splitting surfaces or their microscopic description. In this work, we explore this approach.

In particular, we aim to (i) explore matching scenarios that incorporate cosmological aspects and (ii) investigate if nonsingular systems initially fulfilling the energy conditions can evolve in a way to avoid singularities. We focus on a case where the exterior spacetime is described by the Vaidya metric \citep{Vaidya,Bondi}, representing a radiation-dominated astrophysical system, while the interior spacetime is of the Friedmann-Lema\^itre-Robertson-Walker (FLRW) type, characterized by a scale factor and an arbitrary spatial curvature. We also consider the presence of a scalar field with and without a minimal coupling to the metric. 

Energy conditions play a crucial role in investigating the possibility of singularities in gravitationally collapsing systems. According to Penrose's and Hawking's theorems \cite{Penrose,1970RSPSA.314..529H,haw_ellis_1973}, the existence of trapped surfaces and the fulfilling of the null energy condition (NEC) and the strong energy condition (SEC) are linked to astrophysical and cosmological singularities, respectively. In our analysis, we only focus on situations where the NEC is fulfilled at all times to prevent thermodynamic instabilities but we do not disregard the cases where the SEC could be violated (due to the nonlinearity of the equations). The singularity problem thus becomes nontrivial in our matched spacetimes because we allow for cases where all the energy conditions are met in the exterior spacetime, which might imply the formation of a singularity for external observers, while some of them could be violated in the internal spacetime, naively meaning that singularities could be avoided for internal observers. Clearly, the formation of singularities should be observer-independent. 

We also investigate the role of the coupling between a scalar field and the metric in forming or avoiding singular systems. The motivation for dealing with a scalar field is inspired by the problem of dark energy \citep{1988PhRvD..37.3406R} and inflationary models \cite{1981JETPL..33..532M} near the cosmological singularity, where the violation of the SEC changes the behavior of the scale factor. Due to the intrinsically transient nature of the singularity formation problem when a collapsing system is already compact, it is natural to Taylor expand the scale factor and the scale field around a reference time. The Vaidya spacetime is a natural choice for external observers, particularly in the local universe, as all astrophysical systems emit electromagnetic radiation. Although extreme events like supernovae can deviate from spherical symmetry \citep{2008ARA&A..46..433W}, that should not be that large \citep{2019gwa..book.....A} and assuming a spherically symmetric spacetime for an evolving system (such as a star) is a reasonable approximation. Finally, for a gravitationally collapsing object, due to its intrinsic dynamics, it is reasonable to assume that the internal spacetime is of FLRW-type.

We show that a perturbative analysis of our coupled system of equations has solutions with and without the non-minimal coupling for the first orders in a Taylor-expanded scale factor and scalar field. We investigate the evolution of our systems by forcing them to fulfill all energy conditions and have positive pressure and energy density at the initial (reference) time of collapse. The fulfillment of the energy conditions at later times will be determined by the dynamics of the field equations. We find that the formation of singularities in our matched spacetimes has to respect the energy conditions at all times for the whole spacetime (internal and external to the collapsing object). However, when the SEC is violated--even for an interval of time--in the internal spacetime (and the NEC is always fulfilled), the formation of singularities can be avoided as seen by external observers, even if the energy conditions for them are met. That was found in the case of minimal and non-minimal couplings, although the aspects for so are very different in both cases. The avoidance of singularities stresses the global validity of conditions of the singularity theorems, even for regions of spacetime that are not accessible to observers and times in their past or future. At the same time, the dynamics of the matched spacetimes may offer partial diagnostics to what is hidden from external observers: the bounce is a reflex of the violation of the SEC, which could be due to the presence of a scalar field. 

In Sections \ref{interior} and \ref{exterior}, we present the interior and exterior regions of the spacetime and their respective field equations. In Section \ref{junctions}, the equations obtained after matching both spacetimes through a hypersurface are given. Section \ref{perturbations} is devoted to the self-consistent perturbative treatment of our system of equations, as well as relevant results when the energy conditions are also taken into account. In Section \ref{conclusions} we summarize the main points of this work.

\section{Interior solution}
\label{interior}
As a first step for describing the interior of compact objects, one can take it as homogeneous and isotropic, whose infinitesimal line element in spherical coordinates is given by the FLRW metric 
\begin{equation}
\label{mi1}
ds^2 = -dt^2 + \dfrac{a^2(t)}{b^2(r)}dr^2 + a^2(t)r^2d\theta^2 + a^2(t)r^2\sin^2\theta d\phi^2,
\end{equation}
where $b(r)\equiv \sqrt{1-kr^2}$ and $k$ is a constant representing the spatial curvature. In addition, we assume a non-minimally coupled scalar field $\phi$ as the source for the metric $g_{\mu\nu}$. The total action of the theory is taken as
\begin{equation}
\label{acaoce}
S = \int\sqrt{-g}\,d^4x\left(\dfrac{1}{2}M_{\textrm{pl}}^2\,\mathcal{R}-w-\dfrac{\alpha}{2}\mathcal{R}\phi^2 - V(\phi)\right),
\end{equation}
where $w = (1/2)g^{\mu\nu}\nabla_{\mu}\phi\nabla_{\nu}\phi$ is the kinetic term, $\mathcal{R}$ is the Ricci scalar, $\alpha$ is the coupling parameter, $V(\phi)$ is the potential encoding the possible self-interacting terms of the scalar field and $M_{\textrm{pl}}$ is the Planck mass (with geometric units $c=\hbar=1$). Inspired by the inflationary mechanism \cite{Starobinsky80,Guth81,Starkman}, we adopt a general fourth-order polynomial profile for the potential:
\begin{equation}
\label{potencialce}
V(\phi) = V_0 + \dfrac{\mu^2}{2}\phi^2+\dfrac{\beta}{3}\phi^3+\dfrac{\lambda}{4}\phi^4,
\end{equation}
where $V_0$, $\mu$, $\beta$, and $\lambda$ are constants to be determined by the evolution of the gravitational collapse. 

The equations of motion for $\phi$ and $g_{\mu\nu}$ can be obtained straightforwardly from Equation \eqref{acaoce}. Variation of $S$ with respect to the scalar field yields
\begin{equation}
\label{eqce}
\phi_{tt} + 3H\phi_{t}+V'+\alpha \mathcal{R}\phi = 0,
\end{equation}
where $\phi$ is assumed to be a function of time only. The subscript $t$ means the derivative with respect to the time coordinate and the prime is the derivative with respect to $\phi$. Variation of $S$ with respect to the metric $g_{ab}$ gives the Einstein field equations (EFE) $G_{ab}=(1/M_{\textrm{pl}}^{2})T_{ab}$, with the ``effective'' energy-momentum tensor written as
\begin{equation}
\label{ener_mom_tensor}
T_{ab}= (1 - 2\alpha) \nabla_{a}\phi \nabla_{b}\phi - 2 \alpha\phi\nabla_{a}\nabla_{b}\phi + \alpha G_{ab}\phi^2 - \left[(1 - 4\alpha) w + V(\phi) - 2\alpha\phi\Box \phi\right] g_{ab}.
\end{equation}
The nontrivial components of the EFE are
\begin{equation}
\label{eqc0}
H^2 = \dfrac{\rho}{3M_{\textrm{pl}}^2}-\dfrac{k}{a^2},
\end{equation}
and
\begin{equation}
\label{eqc1}
H_t = -\dfrac{1}{2M_{\textrm{pl}}^2}(\rho + p)+\dfrac{k}{a^2},
\end{equation}
where $H(t)$ is the Hubble parameter defined as $H(t) \equiv a_t(t)/a(t)$, $\rho$ is the energy density and $p$ is the pressure, both calculated in terms of a class of normalized, time-like observers of the form $V^{a}=\delta^{a}_0$. Their explicit expressions are, respectively,
\begin{equation}
\label{rhoce}
\rho = \dfrac{1}{2}\phi_t^2+\dfrac{\alpha}{2}\mathcal{R}\phi^2+V-3\alpha\dfrac{a_{tt}}{a}\phi^2,
\end{equation}
and
\begin{equation}
\label{pce}
p = \dfrac{1}{2}(1-4\alpha)\phi_t^2-\dfrac{\alpha}{2}\mathcal{R}\phi^2-V+\alpha\left[\dfrac{a_{tt}}{a}+2\left(\dfrac{a_{t}}{a}\right)^2+\dfrac{2k}{a^2}\right]\phi^2-2\alpha\phi\phi_{tt},
\end{equation}
with the Ricci scalar here given by 
\begin{equation}
\label{ricci_scalar}
\mathcal{R}=6\left[\dfrac{a_{tt}}{a}+\left(\dfrac{a_{t}}{a}\right)^2+\dfrac{k}{a^2}\right].
\end{equation}
Note that neither $\rho$ nor $p$ have a fixed sign \textit{a priori}. This will lead to a complication in the determination of the initial conditions for the gravitational collapse since they should be positive in the moment the contraction starts. The qualitative analysis of this type of model can be found in \cite{amendola}, where it is claimed that singularity-free solutions are possible for a certain (very limited) class of parameters.

In order to control their behavior closely, it is worth writing the energy conditions explicitly for the above theory. In particular, the NEC is fundamental for the dynamics of the gravitational collapse and is given by
\begin{equation}
\label{gen_nec}
\rho + p = (1-2\alpha)\phi_t^2-2\alpha\left[\dfrac{a_{tt}}{a}-\left(\dfrac{a_{t}}{a}\right)^2-\dfrac{k}{a^2}\right]\phi^2-2\alpha\phi\phi_{tt}\geq 0,
\end{equation}
and it depends explicitly on the spatial curvature. On the other hand, the SEC, responsible for the singular behavior of the interior, is expressed as
\begin{equation}
\label{gen_sec}
\rho +3p = 2(1-3\alpha)\phi_t^2-6\alpha\dfrac{a_{tt}}{a}\phi^2-2V-6\alpha\phi\phi_{tt} \geq 0,
\end{equation}
which depends explicitly on the potential. We must add to them the further hypothesis of the positivity of the energy density ($\rho\geq 0$).

For the sake of completeness, we introduce the mass function \cite{1966ApJ...143..452H,10.1063/1.1665273,PhysRevD.41.3252}, which describes the loss of mass by the compact object in a spherically symmetric collapse: 
\begin{equation}
m(t,r)\equiv \frac{ar}{2}R^{2}{}_{323}=\frac{ar}{2}(r^2a_t^2+kr^2)=\frac{a^3r^3}{6}\rho,\label{mass_function}
\end{equation}
where we used the Friedmann equation (\ref{eqc0}) in the last equality.

\section{Exterior solution}
\label{exterior}
We assume the exterior part of the compact object to be modeled by Vaidya's metric. Written in a radiative coordinate system \cite{Vaidya,Bondi} it takes the form 
\begin{equation}
\label{me0}
ds^2 = -\left(1-2\dfrac{\Bar{m}(u)}{R}\right)du^2+2\varepsilon dudR + R^2\left(d\Bar{\theta}^2+\sin^2\bar{\theta}d\Bar{\phi}^2\right),
\end{equation}
where the $R$-coordinate represents the exterior radius, $u$ is a light-like time coordinate and $\bar m$ is the geometric mass function of the compact object. The free parameter $\varepsilon$ indicates if the radiation is incoming ($\varepsilon=1$) or outgoing ($\varepsilon=-1$). The source for this metric is commonly interpreted as purely incoherent electromagnetic radiation, whose energy-momentum tensor is
\begin{equation}
\label{tmer}
\Bar{T}_{ab} = \dfrac{2\varepsilon}{R^2}\dfrac{d\Bar{m}(u)}{du}l_{a}l_{b},
\end{equation}
where $l_{a}$ is a light-like vector field satisfying $l_{a}dx^{a} = -du$.

Written in this coordinate system, the $g_{00}$ component of the metric plays a crucial role in determining trapped surfaces for the Vaidya's metric \cite{PhysRevD.54.4862}, which are not the event horizon of this spacetime \citep{2004CQGra..21.1135V,doi:10.1142/S0218271811020354}. It is proportional to the product $\bar\chi$ of the expansion factors of each radial null geodesics (incoming and outgoing), which is
\begin{equation}
\label{chibarra}
\Bar{\chi}=1-\frac{2\Bar{m}(u)}{R}.
\end{equation}
When $\bar\chi$ vanishes, a marginally trapped surface is formed and if the radius of the star crosses this boundary, the evolution may eventually lead to a singularity. 

With the introduction of the junction conditions, it is also possible to relate the exterior and the interior radii \cite{Bittencourt2019}. We note that this analysis is of great importance because it is one of the requirements for the validity of the singularities theorems \cite{Penrose}.

\section{Junction conditions}
\label{junctions}
The match of the two aforementioned spacetimes is done by assuming a well-defined 3-dimensional space-like hypersurface $\Sigma$ at the interface described by a coordinate system $\{\xi^{A}\} = \{\tau,\vartheta,\varphi\}$. Each part of the matched spacetimes induces a metric tensor onto $\Sigma$ through the map
\begin{equation}
\label{metind}
    g_{ab}^{\pm} = g_{\mu\nu}^{\pm}\dfrac{\partial x^{\mu}_{\pm}}{\partial \xi^{a}}\dfrac{\partial x^{\nu}_{\pm}}{\partial\xi^{b}},
\end{equation}
such that the infinitesimal line elements on $\Sigma$ become
\begin{equation}
\label{eli}
    ds^2_{\pm} = g_{ab}^{\pm}d\xi^{a}d\xi^{b}.
\end{equation}
Henceforth, the interior solution will be denoted by ``$-$'', while the exterior solution will be denoted by ``+''. 

Applying the first Darmois' junction condition \cite{darmois}, it is guaranteed the continuity of the first fundamental form, namely,
\begin{equation}
\label{first_darmois}
ds^2_{+}\Big|_{\Sigma}=ds^2_{-}\Big|_{\Sigma},
\end{equation}
from which we easily obtain, using Equations (\ref{mi1}) and (\ref{me0}),
\begin{eqnarray}
&&\Dot{t}^2-\dfrac{a^2(t)}{b(r)^2}\Dot{r}^2 = \Bar{\chi}\Dot{u}^2 - 2\varepsilon\Dot{u}\Dot{R},\label{jsuper1}\\[1ex]
&&a(t)r = R,\label{jsuper2}
\end{eqnarray}
where the dot means derivative with respect to $\tau$ (the time coordinate on $\Sigma$). 

The second Darmois' junction condition demands the continuity of the second fundamental form, which is represented by the extrinsic curvature $K_{ab}$, 
\begin{equation}
\label{second_darmois}
K_{ab}^{+}\Big|_{\Sigma} = K_{ab}^{-}\Big|_{\Sigma},
\end{equation}
with
\begin{equation}
\label{curvex}
K_{ab}^{\pm} = -n_{\alpha}^{\pm}\dfrac{\partial}{\partial \xi^{a}}\left(\dfrac{\partial x^{\alpha}_{\pm}}{\partial \xi^{b}}\right)-n_{\alpha}^{\pm}\left(\Gamma^{\alpha}_{\mu\nu}\right)^{\pm}\dfrac{\partial x^{\mu}_{\pm}}{\partial \xi^{a}}\dfrac{\partial x^{\nu}_{\pm}}{\partial \xi^{b}},
\end{equation}
where $n_{\alpha}^{\pm}$ are unit normal vectors orthogonal to $\Sigma$, $(\Gamma^{\alpha}_{\mu\nu})^{\pm}$ are the Christoffel symbols related to each disjoint region of the spacetime. In particular, the corresponding normal vectors are
\begin{equation}
\label{vetnor}
\begin{array}{c}
n_{\mu}^{-} = \dfrac{\epsilon a \Dot{t}}{\sqrt{b^2\Dot{t}^2-a^2\Dot{r}^2}}\left(-\dfrac{\Dot{r}}{\Dot{t}},1,0,0\right), \\ \\
n_{\mu}^{+} = \dfrac{\Bar{\epsilon}}{\sqrt{\Dot{u}(\Bar{\chi}\Dot{u}-2\varepsilon\Dot{R})}}\left(-\Dot{R},\Dot{u},0,0\right),
    \end{array}
\end{equation}
where $(\epsilon,\bar{\epsilon})=\pm 1$ depending on the orientation of $n_{\mu}^{\pm}$. With this in mind, we can calculate explicitly the components of the extrinsic curvature on both sides of $\Sigma$. 

From the junction conditions (\ref{first_darmois}) and (\ref{second_darmois}), after some manipulations, it is possible to write the coordinate system of the exterior spacetime completely in terms of the interior one for this spherically symmetric collapse (for further details, see \cite{PhysRevD.54.4862,Bittencourt2019,fayos2}). Thus, we derive the following set of differential equations completely describing the evolution of the collapse \cite{fayos2}: 
\begin{eqnarray}
&&r_t = \dfrac{\epsilon\Bar{\epsilon}\varepsilon b}{a}\frac{p}{\rho},\label{rponto}\\[1ex]
&&\Bar{m}_t = \dfrac{r^2a^2}{2}\left(\epsilon\Bar{\epsilon}\varepsilon b-ra_t\right)\frac{p}{M_{\textrm{pl}}^2},\label{mponto}\\[1ex]
&&u_t= \frac{1}{\left(\epsilon\Bar{\epsilon}b-\varepsilon ra_t\right)}\frac{\rho+p}{\rho},\label{upt}
\end{eqnarray}
where we used the arbitrariness of the choice of the time coordinate on $\Sigma$ to set $\tau=t$. Equation \eqref{rponto} corresponds to the evolution of the interior radial coordinate; Equation\ (\ref{mponto}) describes the evolution of the mass function; Equation\ (\ref{upt}) gives the time evolution of the exterior in terms of the interior. Note that these expressions are rather general, regardless of the form of $b(r)$, the evolution of the scale factor $a(t)$, or the equation of state.

A solution of the above dynamical system is unique if a consistent set of initial conditions is provided. In order to characterize a gravitational collapse, it is common to demand that at the initial time ($t=t_0$) the following physical conditions \cite{PhysRevD.54.4862,Bittencourt2019} are met:
\begin{equation}
\label{icr}
r_t(t=t_0)<0,
\end{equation}
and also
\begin{equation}
\label{icm}
\bar{m}_t(t=t_0)<0.
\end{equation}
In addition, we must add to our analysis a further equation concerning the exterior radius $R$, since there will be solutions with an increasing scale factor that, according to Equation (\ref{jsuper2}), could lead to an increasing radius for external observers. To avoid such undesired behavior, we also require that 
\begin{equation}
\label{icr_ext}
R_t(t=t_0)<0.
\end{equation}
Again, this condition would be necessary only if $a(t)$ is growing at $t=t_0$. Otherwise, it will be identically satisfied. Finally, from Equations (\ref{first_darmois}) and (\ref{second_darmois}), it is possible to write Eq. (\ref{chibarra}) in terms of the variables of the interior as
\begin{equation}
    \label{chibarra2}
    \Bar{\chi}\Big|_{\Sigma}=b^2-r^2a_t^2,\quad \Longrightarrow \quad \bar m(u)\Big|_{\Sigma}=m(t,r)
\end{equation}
Hence, the apparent 3-horizon occurs when $r a_t = b$. 

It should also be pointed out that this is a simplified model in which possible degrees of freedom on $\Sigma$ are neglected. On the other hand, such simplification guarantees the equivalence of the different junction conditions (see Refs.\ \cite{Mansouri,Lake}). More general cases could also contain surface degrees of freedom \citep{poisson_2004}, where the second fundamental form is discontinuous. However, we ignore this possibility in this work for the sake of simplicity.  

\section{Perturbative analysis}
\label{perturbations}
Given the complexity of the above set of equations, we will approach the problem from a perturbative perspective, using a power law approximation for the scale factor and the scalar field around a reference time (which we take as $t=0$). In particular, we are interested in imposing the energy conditions at this moment to see if a gravitational collapse leads necessarily to a singularity or if it can be avoided by some mechanism due to nonlinearities. The motivation for doing this is to assess the influence of the energy conditions on the singularity theorems. For instance, when the SEC is violated, it is expected a critical point in the scale factor (a bounce). On the other hand, the violation of NEC leads to a mismatch between interior and exterior time coordinates. 

In this vein, we assume a power law expansion of the form
\begin{equation}
\label{fatore}
a(t) = a_b+a_2t^2 + a_4t^4+ \mathcal{O}(t^6),
\end{equation}
and 
\begin{equation}
\label{campoe}
\phi(t) = \phi_0 + \phi_2t^2 + \phi_4 t^4 + \mathcal{O}(t^6),
\end{equation}
where a bounce in the scale factor ($a_b\neq0$) is preferred in order to assess the influence of the energy conditions. Such a hypothesis is reasonable due to the decoupling of the odd and even powers of the expansion when analyzed by the equations of motion of the inner region. Therefore, we neglect the odd powers from the beginning. 

For completeness, we will inspect the effect of each sign of the spatial curvature separately. Thus, for different values of $k$, one can integrate Equation (\ref{rponto}) and obtain 
\begin{equation}
\label{radius}
r(t)=\left\{\begin{array}{l}
r_0+\displaystyle{\int}_{\hspace{-0.2cm}t_0}^{t}\dfrac{\epsilon\Bar{\epsilon}\varepsilon }{a}\fracc{p}{\rho}\,dt,\qquad \mbox{for}\quad k=0,\\[3ex]
r_0\cos\left(\sqrt{k}\displaystyle{\int}_{\hspace{-0.2cm}t_0}^{t}\dfrac{\epsilon\Bar{\epsilon}\varepsilon }{a}\fracc{p}{\rho}\,dt\right)+\sqrt{\frac{1}{k}-r_0^2}\,\sin\left(\sqrt{k}\displaystyle{\int}_{\hspace{-0.2cm}t_0}^{t}\dfrac{\epsilon\Bar{\epsilon}\varepsilon }{a}\fracc{p}{\rho}\,dt\right),\qquad \mbox{for}\quad k>0\\[3ex]
r_0\cosh\left(\sqrt{|k|}\displaystyle{\int}_{\hspace{-0.2cm}t_0}^{t}\dfrac{\epsilon\Bar{\epsilon}\varepsilon }{a}\fracc{p}{\rho}\,dt\right)+\sqrt{\frac{1}{\sqrt{|k|}}+r_0^2}\,\sinh\left(\sqrt{|k|}\displaystyle{\int}_{\hspace{-0.2cm}t_0}^{t}\dfrac{\epsilon\Bar{\epsilon}\varepsilon }{a}\fracc{p}{\rho}\,dt\right),\qquad \mbox{for}\quad k<0,
\end{array}\right.
\end{equation}
where the initial condition for $r(t)$ set as $r(t_0)=r_0$, and $t=t_0$ indicates the beginning of the collapse. As long as $r(t)$ is determined, the equations for $\Bar{m}(t)$ and $u(t)$ can be found by quadrature. As one may notice, these expressions will lead to completely distinct behavior for the radius of the compact object.

The first case we briefly discuss here has no physical interest. It is obtained by inserting the Taylor expansions (\ref{fatore}) and (\ref{campoe}) into the equations of motion for the interior (\ref{eqce}), (\ref{eqc0}), and (\ref{eqc1}) with $k=0$. The outcome is a set of equations that are manifestly incompatible with respect to the free parameters. Although there are more parameters than equations, the first orders lead solely to the trivial result: $a_2=a_4=\phi_2=\phi_4=0$, namely, the scale factor and the scalar field must be constant. We have also checked that this still remains for higher orders. Indeed, we verified this up to $\mathcal{O}(t^8)$. From this, we conclude that solutions for the interior might be possible only if very high orders are taken into account or for functions that do not admit a Taylor expansion around the origin. Both options are out of the scope of this work.

For the cases with a nonzero spatial curvature, we assume the same Taylor expansion given by Equations (\ref{fatore}) and (\ref{campoe}). We insert them into the field equations (\ref{eqce}), (\ref{eqc0}) and (\ref{eqc1}) and solve order by order. As a result, the problem is reduced to a set of nine coupled equations. The unknowns are $a_b$, $a_2$, $a_4$, $\phi_0$ $\phi_2$, $\phi_4$, $V_0$ $m$, $\beta$, $\lambda$, and $k$, including the non-minimal coupling parameter $\alpha$. Thus, we have 12 constants for 9 equations, meaning that one can choose 3 unconstrained parameters. For convenience, we choose the free parameters as $\phi_0$, $\phi_2$, and $a_b$, where the choice of the first one can be freely made without physical implications, as we will see\footnote{From physical arguments, the appropriate choice should be with $V_0$, instead of $\phi_2$, due to the freedom in the determination of the former. However, this would lead to a sixth-order polynomial for $\phi_2$ which would render our analysis unnecessarily nonanalytical.}. 

From these equations, it is possible to obtain 3 analytic solutions. The first one, corresponding to a minimal coupling solution, is
\begin{equation}
\label{sol1}
\begin{array}{ccl}
\fracc{V_0}{M_{\textrm{pl}}^2} &=& -\fracc{[k^6 -50 k^4 \phi_2^2 + 865 k^2 \phi_2^4 + 192 \phi_2^6]\phi_0^3 + 72( k^2 +11 \phi_2^2) k \phi_2^3 \phi_0^2-864 \phi_0 \phi_2^6-3456 k \phi_2^5}{1152 \phi_2^5},\\ [2ex]
\mu^2 &=& -\fracc{(k^6 \phi_0^2-50 k^4 \phi_0^2 \phi_2^2+865 k^2 \phi_0^2 \phi_2^4+192 \phi_0^2 \phi_2^6+48 k^3 \phi_0 \phi_2^3+528 k \phi_0 \phi_2^5-288 \phi_2^6)}{96 \phi_2^5 \phi_0} ,\\ [2ex]
\beta &=& -\fracc{(k^6 \phi_0^2-50 k^4 \phi_0^2 \phi_2^2+865 k^2 \phi_0^2 \phi_2^4+192 \phi_0^2 \phi_2^6+36 k^3 \phi_0 \phi_2^3+396 k \phi_0 \phi_2^5-144 \phi_2^6)}{48 M_{\textrm{pl}} \phi_2^5 \phi_0^2},\\[2ex]
\lambda &=& \fracc{(k^6 \phi_0^2-50 k^4 \phi_0^2 \phi_2^2+865 k^2 \phi_0^2 \phi_2^4+192 \phi_0^2 \phi_2^6+24 k^3 \phi_0 \phi_2^3+264 k \phi_0 \phi_2^5-96 \phi_2^6)}{96 M_{\textrm{pl}}^2 \phi_2^5 \phi_0^3},\\ [2ex]
a_2 &=& \fracc{k a_b}{2},\quad a_4 = \fracc{(k^2-\phi_2^2)a_b}{24},\quad \phi_4 = \fracc{(k^2-13\phi_2^2) k M_{\textrm{pl}}}{96 \phi_2},
\end{array}
\end{equation}
where we have rescaled the free parameters in order to simplify the expressions as follows
\begin{equation}
k\longrightarrow a_b^2\,k,\qquad \phi_0 \longrightarrow M_{\textrm{pl}}\,\phi_0, \qquad \phi_2 \longrightarrow \frac{1}{2}\,M_{\textrm{pl}}\,\phi_2.
\end{equation}
The range of applicability of this solution is dictated by the inequalities (necessary to guarantee the meaningfulness of the Taylor expansions)
\begin{equation}
\label{ineq_min}
\begin{array}{lcl}
|a_4t^2|\ll |a_2|&\Longrightarrow&\frac{1}{12|k|}|k^2-\phi_2^2| t^2\ll1\\[1ex]
|\phi_4t^2|\ll |\phi_2|&\Longrightarrow &\frac{1}{48|k|\phi_2^{2}}|k^2-13\phi_2^{2}| t^2\ll1.
\end{array}
\end{equation}
The above means that for small $t$ we should also have $\phi_2$ as close as possible to the straight lines $\phi_2^{(1)}=\pm|k|$ and $\phi_2^{(2)}=\pm|k|/\sqrt{13}$. As we will show later, these inequalities are crucial for the choice of viable initial conditions.

Associated with the non-minimal coupling $\alpha=M_{\textrm{pl}}^2/\phi_0^2$, the non-vanishing parameters of the remaining pair of solutions are
\begin{eqnarray}
\label{sol2}
V_0 = \frac{1}{12}\frac{[3\lambda a_b^2 \phi_0^4 + 2\beta a_b^2 \phi_0^3 + \frac{36}{19} k M_{\textrm{pl}}^2 (11\pm3\sqrt{5}) + 36 k M_{\textrm{pl}}^2]}{a_b^2},\quad a_2 = \frac{k(11\pm 3\sqrt{5})}{38\,a_b}\nonumber\\[1ex] 
\mu^2 = -\frac{[\lambda a_b^2 \phi_0^4 + \beta a_b^2 \phi_0^3 + \frac{6}{19} k M_{\textrm{pl}}^2(11\pm3\sqrt{5}) + 6 k M_{\textrm{pl}}^2]}{a_b^2 \phi_0^2},\quad a_4= \frac{k^2(63 \pm 12\sqrt{5})}{2166\,a_b^3},
\end{eqnarray}
with arbitrary $\beta$, $\lambda$, $a_b$, $\phi_0$, and $k$. For clarity, we identify each one of these solutions as the $(\pm)$-solutions. Again, we used the map $k \longrightarrow a_{b}^2 k$ in order to simplify the above expressions. Since $\phi_2=\phi_4 = 0$, the scalar field must be constant 
(ground state) up to this order of approximation. For this pair of non-minimal coupling solutions, the sign of $a_2$ depends only on $k$, since $a_b>0$. We also obtain the range of applicability of the approximations for Equations (\ref{fatore}) and (\ref{campoe}) in this non-minimal case by imposing (see $a_2$ and $a_4$ in Equation \eqref{sol2})
\begin{equation}
|a_4t^2|\ll |a_2|\quad \Longrightarrow \quad |t|\ll\frac{1}{\sqrt{|k|}}\sqrt{\frac{57(11\pm 3\sqrt{5})}{63 \pm 12\sqrt{5}}}=
\left\{\begin{array}{l}
3.35/\sqrt{|k|}\quad \mbox{for the (+)-solution},\\[1ex]
2.60/\sqrt{|k|}\quad \mbox{for the (-)-solution}.      
\end{array}\right.
\end{equation}
It should be remarked that when choosing $\phi_0=\sqrt{6}M_{\textrm{pl}}$, we recover the conformal coupling $\alpha=1/6$. Even if the $(\pm)$-solutions seem very similar, we will see later on that the $(-)$-solution is too restrictive with respect to the validity of the approximations. As we will see, both present an interval in which the SEC condition is violated but the NEC is always satisfied. Henceforth, time is measured in terms of the inverse square root of the spatial curvature and the scalar field is measured in Planck mass units.

\subsection{The minimal coupling solution}
Let us analyze in detail the case corresponding to the minimal coupling solution given by Equations (\ref{sol2}). Although the expressions for the parameters are complicated, the energy density and the pressure are rather simple:
\begin{equation}
\label{rho_p_sol1}
\frac{\rho(t)}{M_{\textrm{pl}}^2}=\frac{3 k(4-\phi_2^2t^4)}{4}\,,\quad\mbox{and}\quad \frac{p(t)}{M_{\textrm{pl}}^2}=\frac{k(k^2 - 4 \phi_2^2) t^4 + 12 t^2\phi_2^2 - 36 k}{12}.
\end{equation}
Note that they are independent of $a_b$ and $\phi_0$. The expressions for the energy conditions can be put in the form
\begin{eqnarray}
k(k^2-13\phi_2^2) t^2 + 12\phi_2^2\geq0,&\qquad& \mbox{for the NEC},\label{NEC_sol1}\\[1ex]
k(k^2-7\phi_2^2) t^4 + 12\phi_2^2 t^2 -24 k\geq0,&\qquad& \mbox{for the SEC}.\label{SEC_sol1}
\end{eqnarray}
It is evident the even character of these functions under the symmetry $t\rightarrow-t$ and $\phi_2\rightarrow-\phi_2$. That is important for determining whether the energy conditions are violated only for an interval of time. Without any loss of generality, one can set $k=\pm1$ by multiplying the scale factor by a constant. Therefore, the different values of $k$ with the same sign will lead to the same qualitative behavior. Under this choice, of course, the magnitude of $a_b$ cannot be chosen arbitrarily anymore.

\begin{figure}[ht]
\centering
    \includegraphics[scale=0.36]{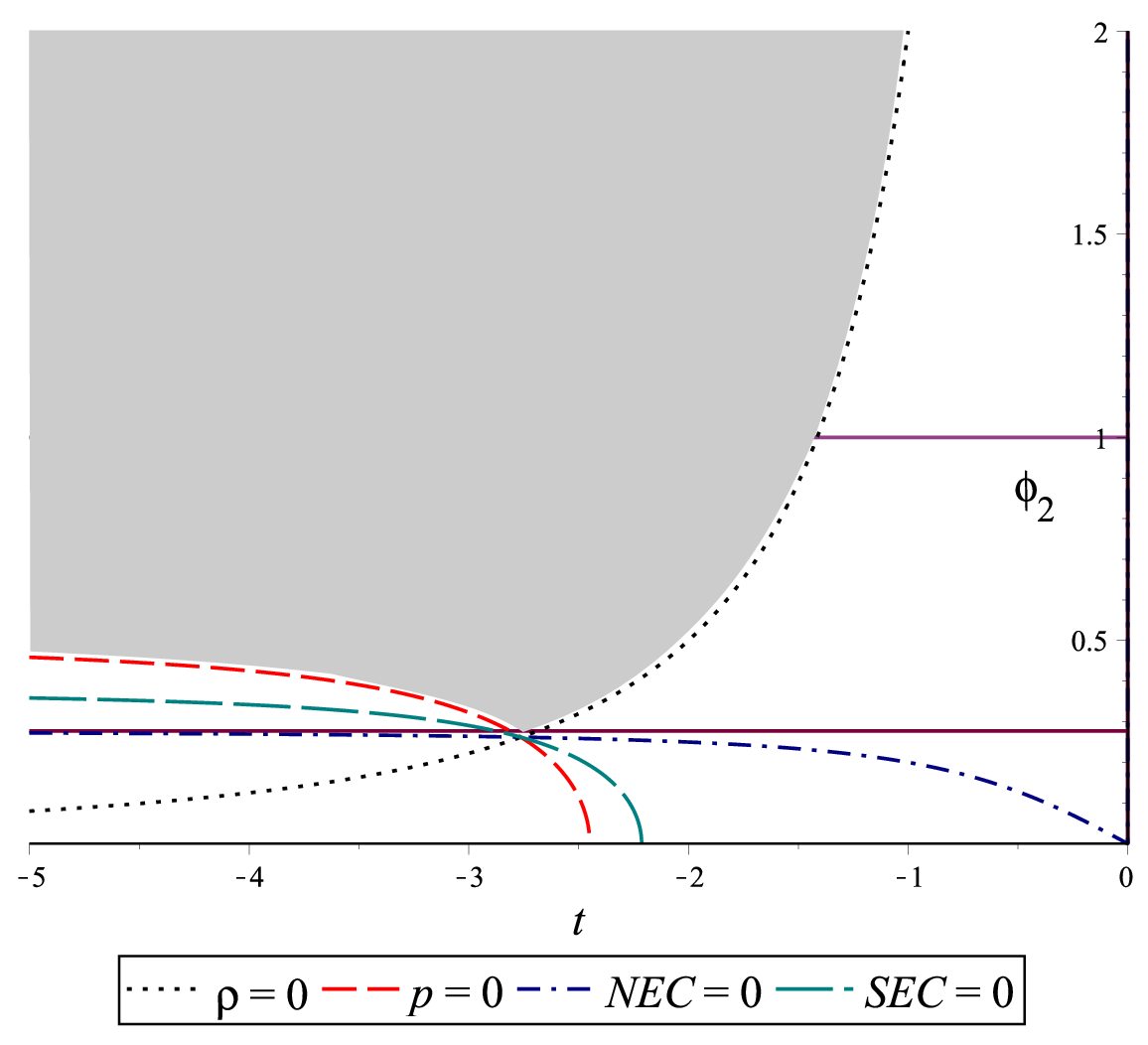}
    \includegraphics[scale=0.36]{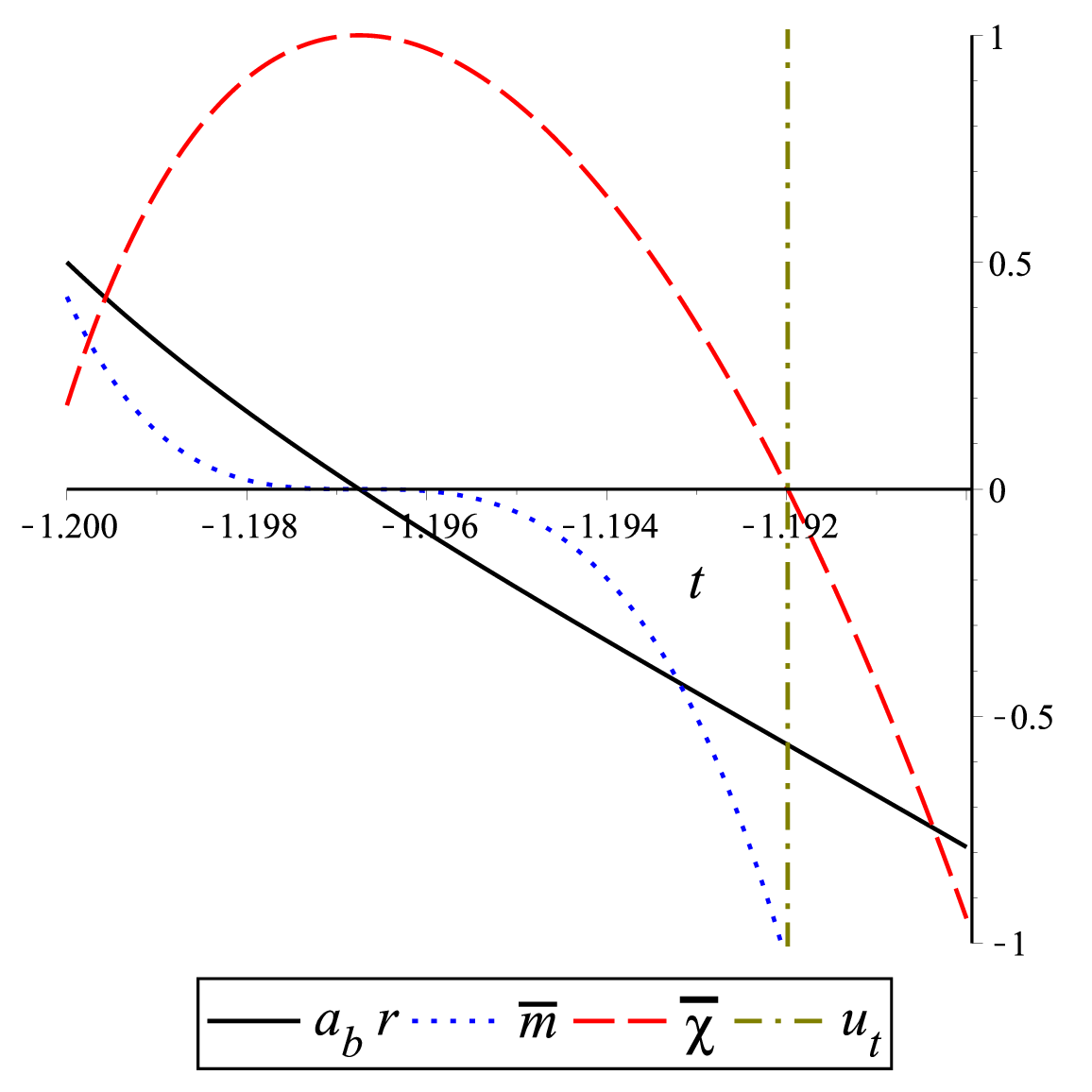}
    \caption{Null contours in the $\phi_2-t$ plane for the energy density, pressure, NEC, and SEC assuming the minimal coupling. On the left panel, the straight lines represent the values of $\phi_2$ which automatically satisfy the Taylor-expansion constraints given by Equation \eqref{ineq_min} for the case $k=-1$. The shaded region corresponds to the set of parameters where $(\rho, p, \mbox{NEC}, \mbox{SEC})>0$, and hence are mathematically and physically allowed initial conditions for the gravitational collapse. As time evolves towards $t=0$ along a given $\phi_2$ within the shaded region, one sees that the energy conditions are never violated. However, for each choice of $\phi_2$, there is a moment when $\rho=0$, meaning that the mass function is null (see Equation \eqref{mass_function}). Thus, depending on when the star radius crosses the apparent 3-horizon ($\bar{\chi}=0$), an evaporation or a singularity are expected for this case. On the right panel, for $\phi_2=2$, $t_0=-1.2$, and $r_0=0.5/a_b$, one sees that a complete evaporation happens. (We multiply the mass function by a factor of 100 to render it visible.) }
    \label{min_curve_neg_k}
\end{figure}

For $k=-1$, it is possible to find proper initial conditions where $\rho$ and $p$ are positive and the energy conditions (NEC and SEC) are fulfilled. This is indicated in Figure \ref{min_curve_neg_k} by the shaded region. The two horizontal lines are those obtained from the inequalities (\ref{ineq_min}). They are not mandatory, since the magnitude of $t$ is also relevant, but the initial conditions cannot be set too far from them. We see the curves $\rho=0$ and $p=0$ and the ones associated with the lower bound of the energy conditions, i.e., NEC=0 and SEC=0, intercept them all at a certain instant of time ($t_c\sim -2.7$ in Figure \ref{min_curve_neg_k}). The evolution of the gravitational collapse towards $t=0$ will inevitably lead the system to cross the curve $\rho=0$ (when $\bar{m}=0$, see Equation \eqref{mass_function}) if the initial mass is not too small, meaning that an evaporation can happen in this case and the Vaidya spacetime becomes Minkowski. From Equation (\ref{rponto}), it implies that the derivative of the radius will negatively diverge and therefore the radius will decrease so fast that a singularity could also be formed. Indeed, when $r(t)$ vanishes, the derivative of the mass function goes to zero (see Equation \eqref{mponto}), meaning that the compact object can be left with a constant mass. However, if we add the requirement that $\bar\chi>0$ at the initial time of collapse, i.e., no trapped surfaces in the beginning, the window of the physical initial conditions is narrowed. In Figure \ref{min_curve_neg_k}, for the case  $\phi_2=2$, $t_0=-1.2$ and $r_0=0.5/a_b$, although the energy conditions are met, a trapped surface is not developed on time and the star evaporates. This example is a clear illustration of the need to fulfill all conditions of the singularity theorems so singularities emerge: the energy conditions are never violated and a trapped surface is present. 

\begin{figure}[ht]
\centering
    \includegraphics[scale=0.25]{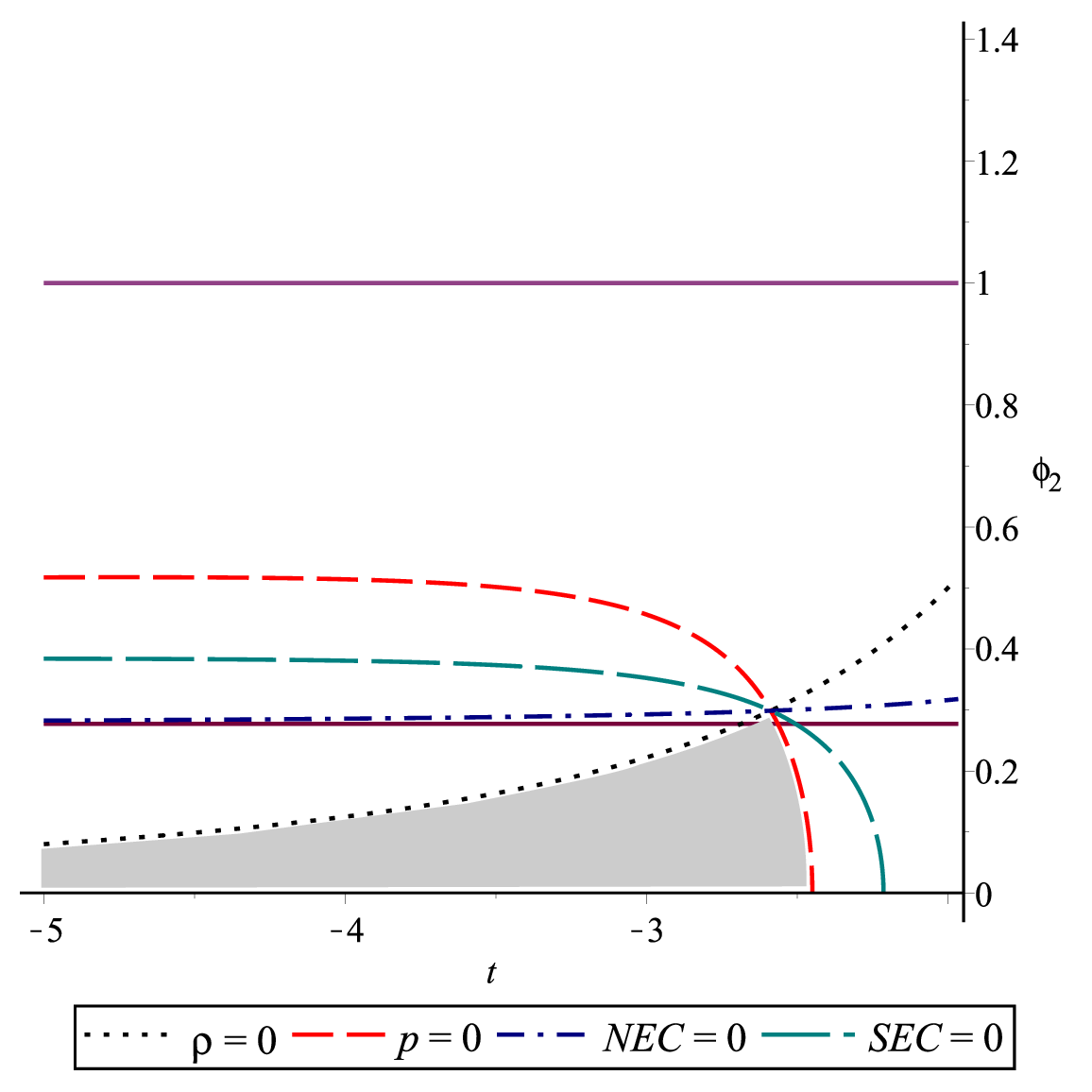}
    \includegraphics[scale=0.25]{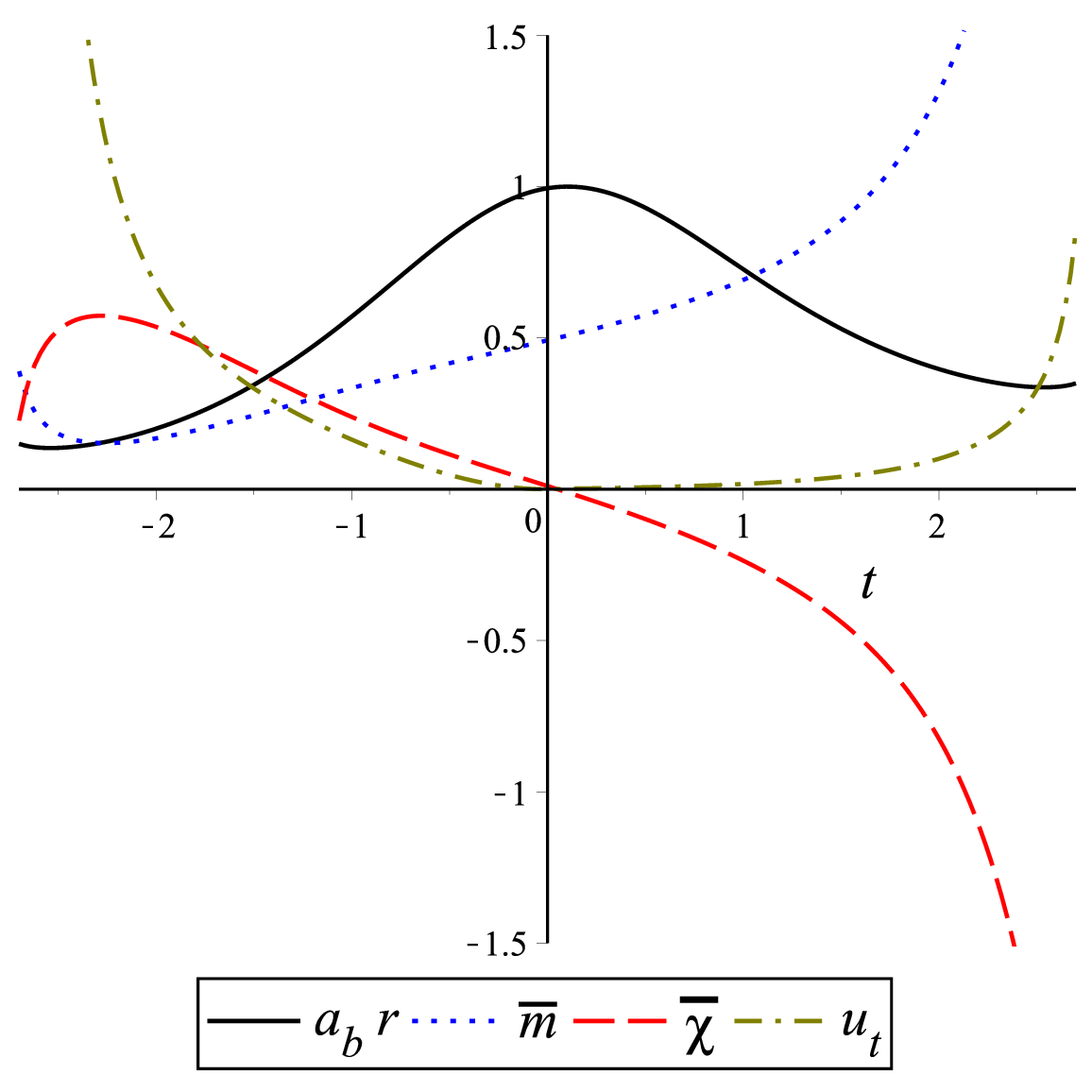}
    \includegraphics[scale=0.25]{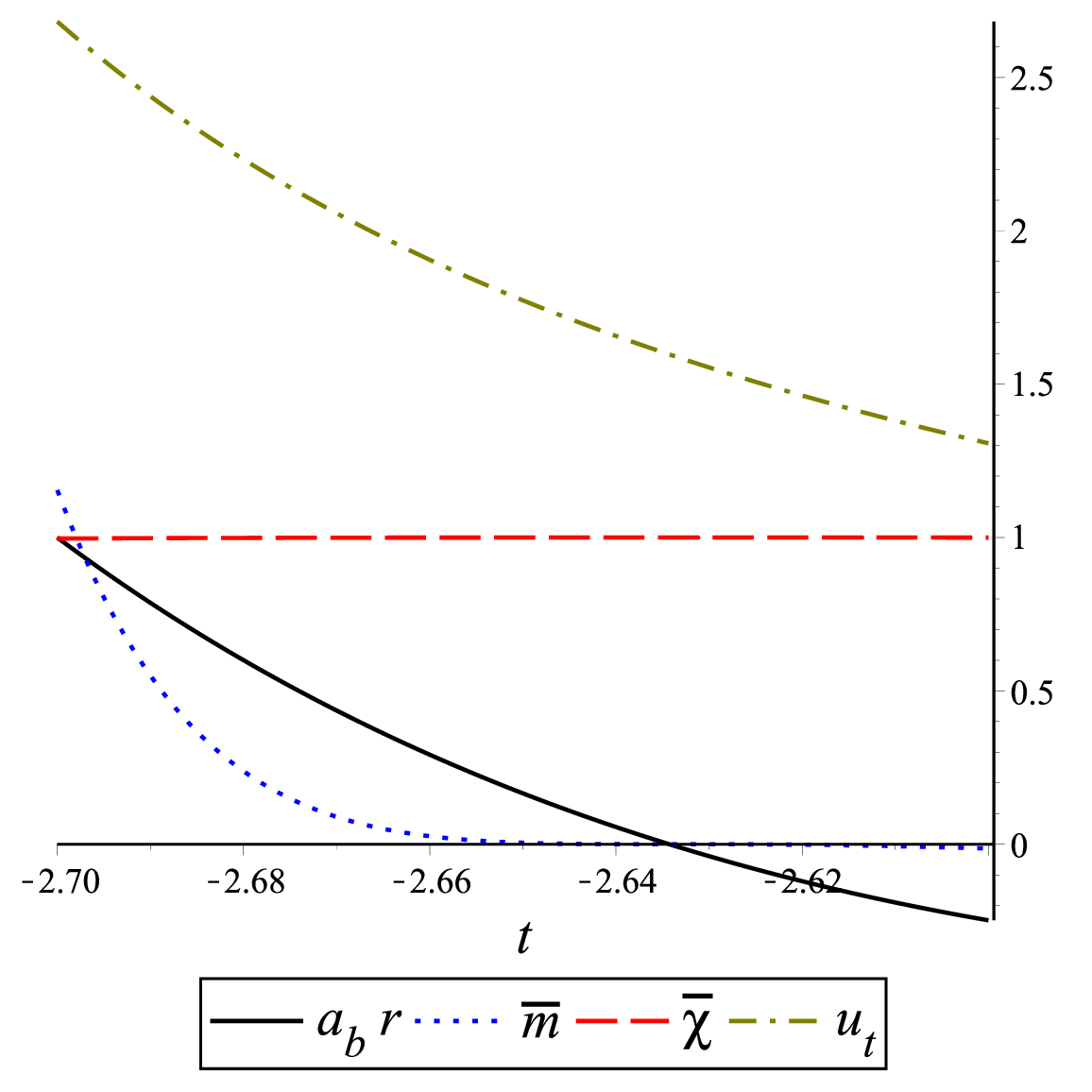}
    \caption{Left: Curves with null pressure, energy density, SEC and NEC for the case $k=1$ and the minimal coupling. The meaning of the shaded area and the straight lines is exactly the same as in Figure \ref{min_curve_neg_k}. Any value of $\phi_2$ close to the straight line and $t_0$ will be such that the evolution of the system towards $t=0$ will make it cross the curve $p=0$ and a bit later, violating the SEC. However, due to its symmetry with respect to $t$, it will be valid again for positive times. As a result, the formation of a singularity is not guaranteed anymore. Middle: For the parameter choices $\phi_2=0.25$, $t_0=-2.7$, and $r_0=0.15/a_b$ we see that the evolution of $r(t)$, $\bar{m}_t$, $\bar{\chi}$ and $u_t$ suggest, however, that a singularity will be formed at later times due to the formation of a trapped surface at $t=0$. Even though the SEC is violated for  $-2.5\lesssim t\lesssim 2.5$, which leads the radius of the star to increase up to a maximum, that is not enough to avoid--but just stall--the singularity. Even though a bounce takes place ($a(t)$ increases), a contraction of the star's radius ($a r\sim a_br$) is not avoided at later times. Right: For the same choice of parameter, but starting with $r_0=0.01/a_b$, the final stage is the complete evaporation of the compact object. (We multiply the radius by a factor of 100 and the mass function by a factor of $10^4$ to render them visible.)}
    \label{min_curve_pos_k}
\end{figure}

\begin{figure}[ht]
\centering
    \includegraphics[scale=0.25]{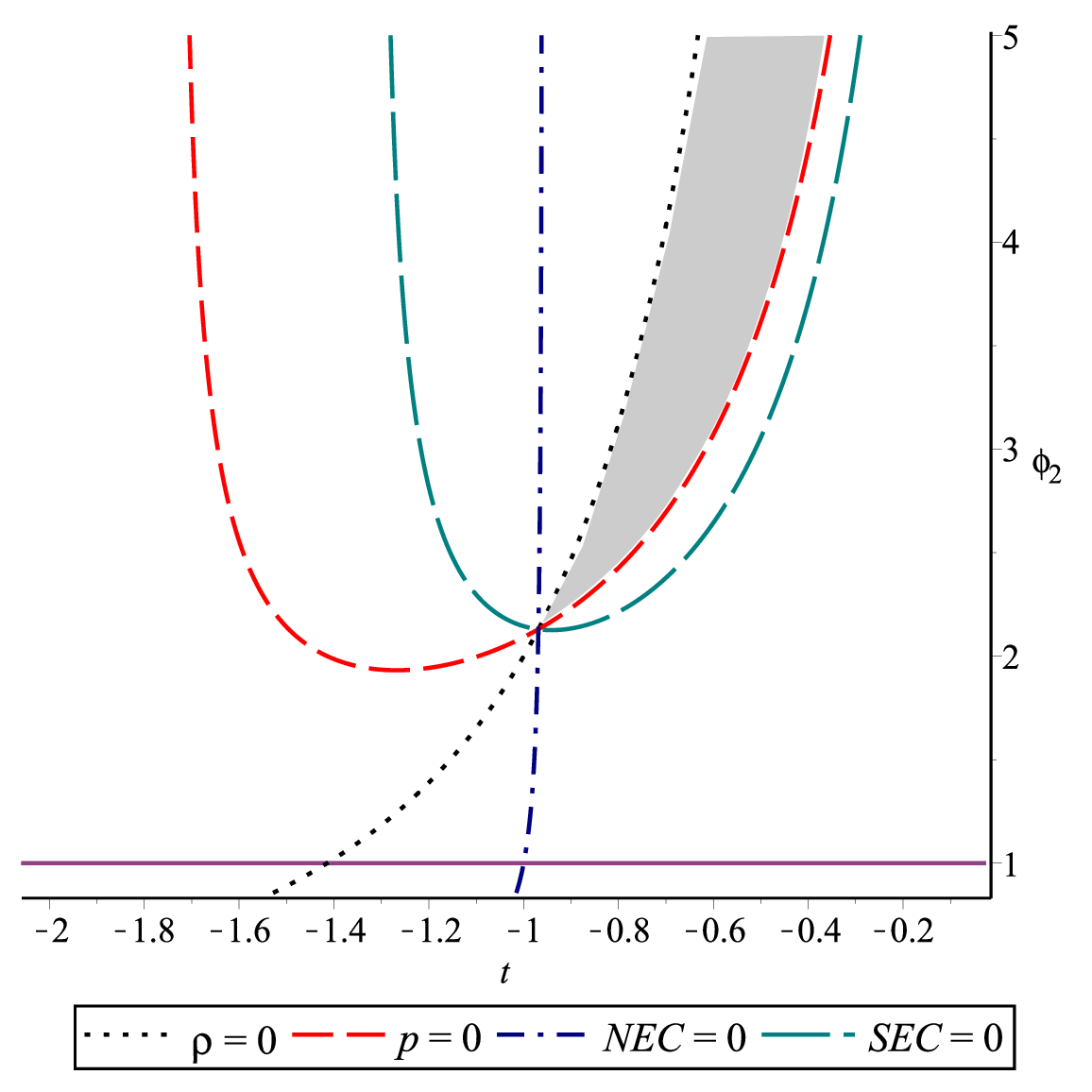}
    \includegraphics[scale=0.25]{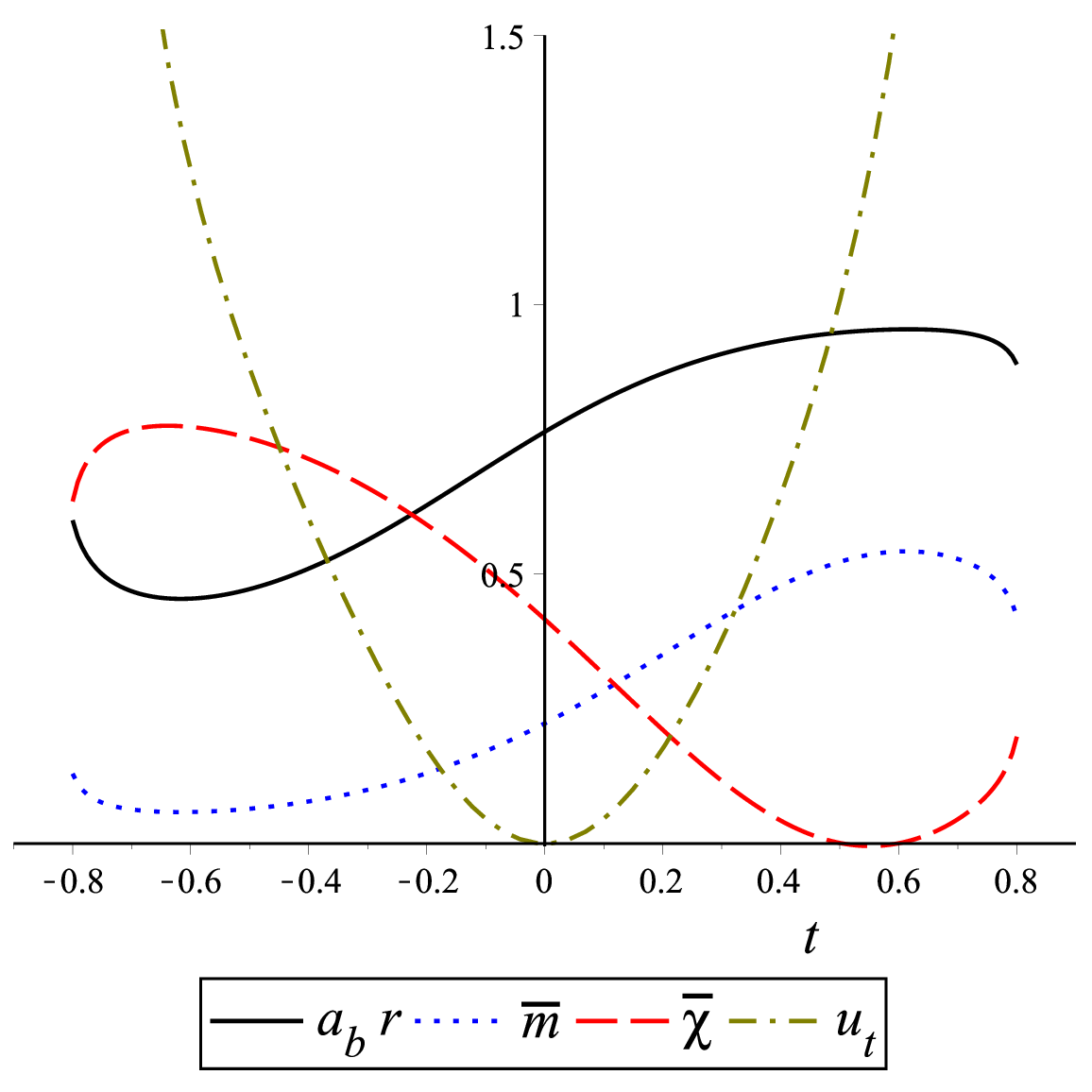}
    \includegraphics[scale=0.25]{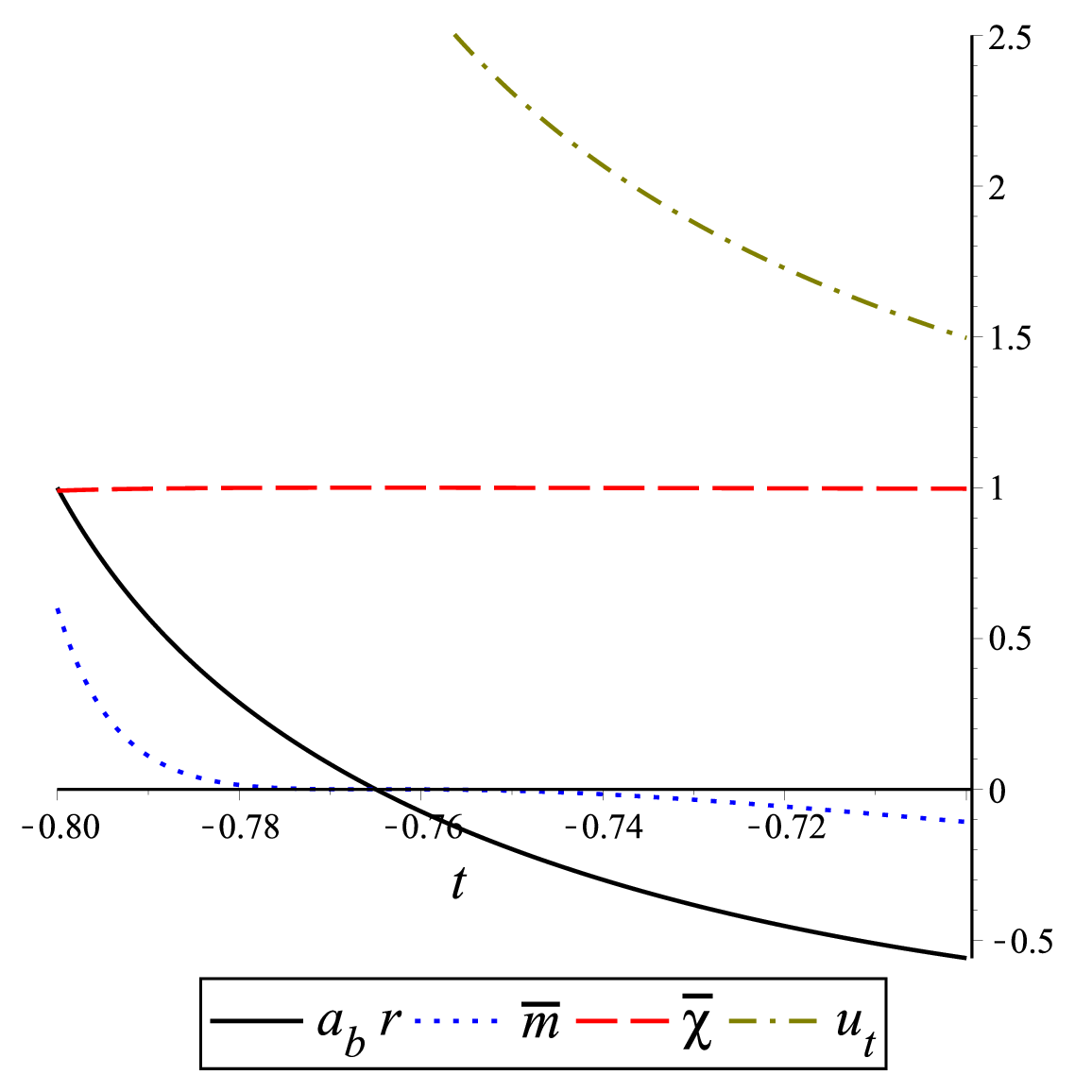}
    \caption{Left: Curves with null pressure, energy density, SEC and NEC for the case $k=1$, and the minimal coupling. The shaded area and the straight lines indicate another region of viable physical initial conditions. Middle: The behavior of $r(t)$, $\bar{m}_t$ and $u_t$ for the parameter choices $\phi_2=3.0$, $t_0=-0.8$, and $r=0.6/a_b$. For the moment the SEC is violated, the star experiences a bounce ($a_br$ increases), but there is a short interval of time where $\Bar{\chi} <0$. Then, $\Bar{\chi} >0$ again. This example shows that a violation of the SEC renders the formation of singularities unpredictable and that trapped surfaces could appear and disappear when that happens. In this case, the collapsing system presents a nonsingular behavior. Right: The behavior of $r(t)$, $\bar{m}_t$ and $u_t$ for the parameter choices $\phi_2=3.0$, $t_0=-0.8$, and $r=0.1/a_b$. For this choice of parameters, evaporation takes place and the Vaidya spacetime becomes Minkowski. (We multiply the radius by a factor of 10 and the mass function by a factor of 1000 to render them visible.)}
    \label{min_pos_k_r_m}
\end{figure}

For the case $k=1$, the qualitative behavior is very different from $k=-1$. The viable initial conditions are indicated by the two shaded regions in the left panels of Figures \ref{min_curve_pos_k} and \ref{min_pos_k_r_m}. Now, we see that it is possible to have a gravitational collapse without crossing $\rho=0$, but, eventually, the pressure will become negative and the SEC will be violated for an interval of time. For this physically acceptable class of initial conditions, the NEC is always satisfied. This is a very interesting case because the outcome of the collapse cannot be determined a priori anymore. In fact, we present two cases to illustrate that, shown in the middle panels of Figures \ref{min_curve_pos_k} and \ref{min_pos_k_r_m}. The first case is obtained by choosing $\phi_2=0.25$ and $t_0=-2.7$ (shaded area of Figure \ref{min_curve_pos_k}). The second one is given by $\phi_2=3$ and $t_0=-0.8$ (shaded area of Figure \ref{min_pos_k_r_m}). For the first choice of initial conditions, the initial radius is $r_0=0.15/a_b$ and $r$ reaches the maximum $r=1/a_b$ at the same that $\bar\chi$ becomes negative (see the middle panel of Fig. \ref{min_curve_pos_k}). After that, the radius starts to decrease while the mass function continues to increase, the SEC becomes valid again and, if there is no mechanism to stop such behavior, the gravitational collapse will form a singularity (out of the domain of validity of our approximations, of course). For the second one, the evolution is smoother: there is a bounce of the star radius, and a trapped surface is only formed momentarily (see the middle panel of Fig. \ref{min_pos_k_r_m}). Its disappearance at later times and the fact that the star is left with a finite mass and radius at the end of the interval of validity of our solution suggests that a singularity behavior has been avoided in this case. This behavior is due to the temporary violation of the SEC. Again, it shows that when some of the conditions of the singularity theorems are violated, even for small intervals of time, the collapse of gravitational systems becomes unpredictable in general. In the right panels of Figures \ref{min_curve_pos_k} and \ref{min_pos_k_r_m}, we illustrate cases of evaporation, assuming compact objects with very low initial radii. Therefore, there are always two ways to avoid singularities as long as our approximations are valid.

\subsection{The non-minimal coupling solution}
We now examine the gravitational collapse in terms of the solutions (\ref{sol2}). First, we evaluate the energy density and the pressure for this solution, which are 
\begin{equation}
\label{rho_sol2}
\frac{\rho_{\pm}(t)}{M_{\textrm{pl}}^2}=\pm\frac{3 k}{13718}\left[\left(137\sqrt{5}\pm 629\right)k^2 t^4+(342\sqrt{5}\mp1634)k t^2\right]+3 k,
\end{equation}
and
\begin{equation}
    \label{p_sol2}
    \frac{p_{\pm}(t)}{M_{\textrm{pl}}^2}=\pm\frac{k}{13718}\left[\left(137\sqrt{5}\pm629\right)k^2 t^4+\left(342\sqrt{5}\mp1634\right)k t^2\right]+\frac{k}{19}\left(\mp 41-6\sqrt{5}\right),
\end{equation}
where the symbol ``$\pm$'' on the left-hand side of Equations (\ref{rho_sol2}) and (\ref{p_sol2}) indicates which sign from the solutions (\ref{sol2}) is taken. From these expressions, we are able to compute and analyze the validity of the energy conditions. For the NEC, we have
\begin{equation}
    \label{NEC_sol2}
    \pm \left[\left(137\sqrt{5}\pm 629\right)k^3 t^4+\left(342 \sqrt{5} \mp 1634\right)k^2 t^2\right]\pm\frac{2 k}{19}\left(-3\sqrt{5} \pm 8\right)\geq0,
\end{equation}
while for the SEC, it follows that
\begin{equation}
    \label{SEC_sol2}
\pm \frac{3}{6859}\left[\left(137 \sqrt{5}\pm 629\right) k^3 t^4+\left(342\sqrt{5}\mp 1634\right)k^2 t^2\right]\pm\frac{3 k}{19}\left(-6\sqrt{5}\mp22\right)\geq0.    
\end{equation}
The complete characterization of the gravitational collapse can be done when we write Equations (\ref{rponto}) and (\ref{mponto}) in terms of the solutions (\ref{sol2}). 

\begin{figure}[ht]
\centering
    \includegraphics[scale=0.25]{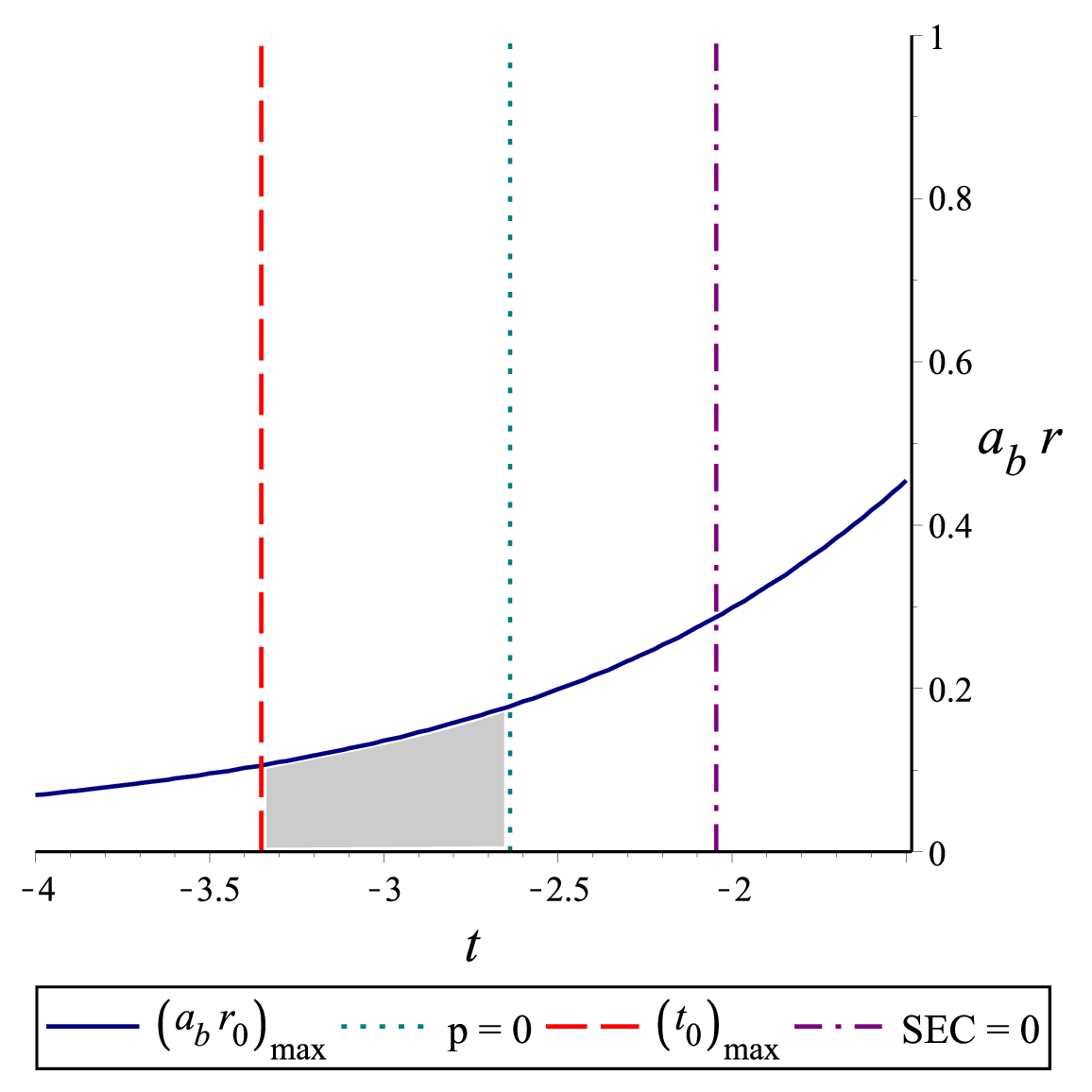}
    \includegraphics[scale=0.25]{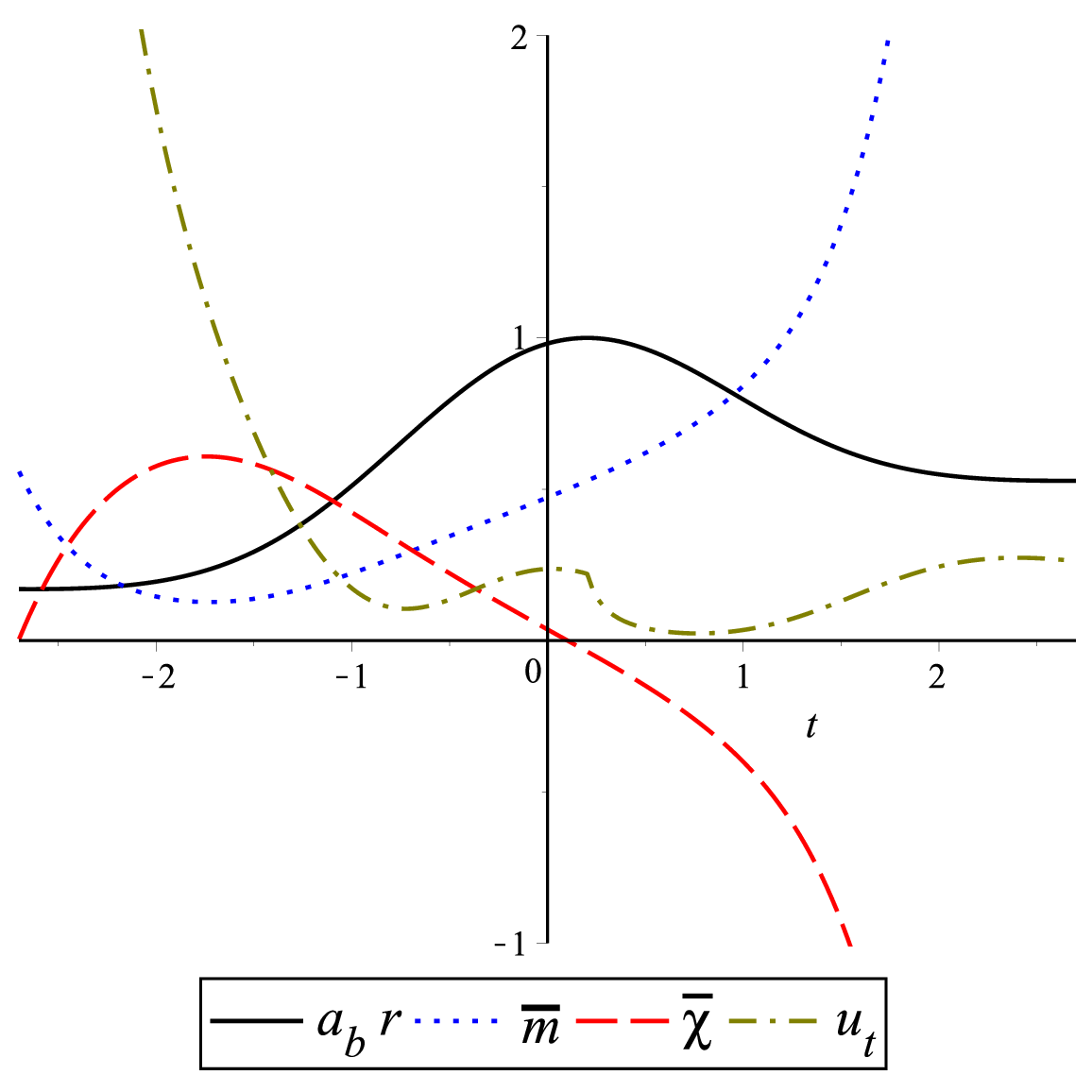}
    \includegraphics[scale=0.25]{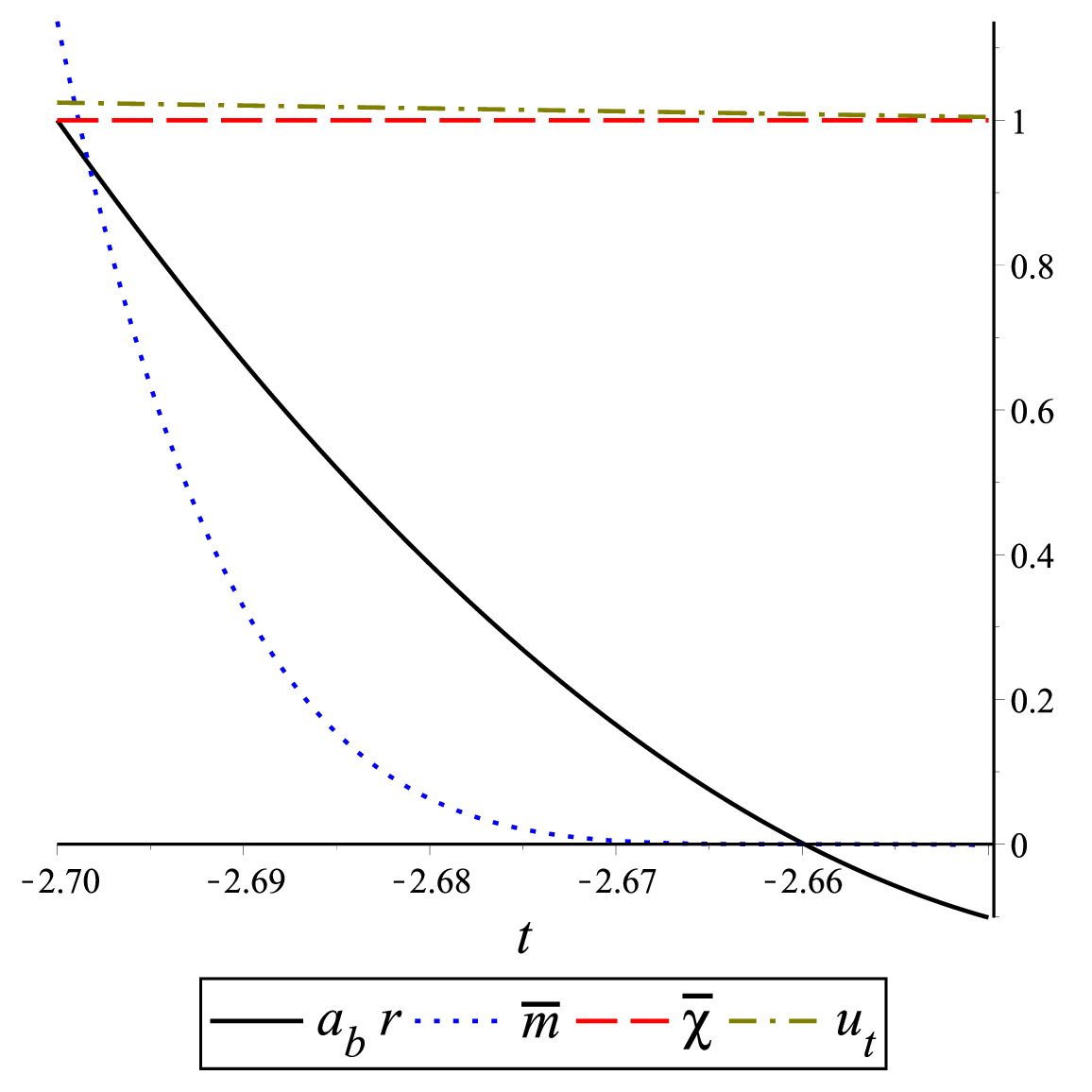}
    \caption{Left: Below the solid curve is the region (shaded) of possible values for $r_0$ and $t_0$ for the $(+)$-solution in the case $k=1$. The dashed (red) vertical line indicates the upper limit of the power law expansion. On the left of the dotted green line and the dot-dashed purple lines, the pressure is positive and the SEC is valid, respectively. When $t\rightarrow 0$, the SEC is violated. The violation takes place up to $t\sim 2$. Middle: However, a singularity is likely to happen for the choice of parameters $t_0=-2.7$ and $r_0 =0.17/a_b$ because a trapped surface appears for $t>0$. The transient violation of the SEC is not enough to avoid the formation of a singularity even though the radius of the star experiences a bounce. Right: Notwithstanding, for certain choices of the parameters, such as $r_0 =10^{-4}/a_b$, a singularity is avoided due to mass evaporation. (We multiply the radius by a factor of $10^{4}$ and the mass function by a factor of $10^{10}$ to render them visible.)}
    \label{initial_conditions_sol2}
\end{figure}

For $k=1$, one can see in the left panel of Figure \ref{initial_conditions_sol2} that the upper limit of the power-law expansion is given by the dashed vertical line in the case of $(+)$-solution. The other vertical lines indicate where the pressure becomes negative (dotted line) and the SEC ceases to be valid (dot-dashed line). (We note that for a symmetric positive time, the SEC becomes valid again.) Below the solid curve are the possible initial radii obtained from Eqs. (\ref{icr}) and (\ref{icm}). In particular, for the choice $t_0=-2.7$ and $r_0 =0.17/a_b$, the evolution of the gravitational collapse is shown in the central panel of Figure \ref{initial_conditions_sol2}. Note that the qualitative behavior of $a_br$ is similar to one in the middle panel of Figure \ref{min_pos_k_r_m}, namely, there is a bounce in the star radius. However, a bit after $t=0$, $\bar{\chi}<0$ (which contrasts with the case in the middle panel of Fig. \ref{min_pos_k_r_m}, where $\Bar{\chi}<0$ only for a small interval of time), meaning that a trapped surface is formed even during the interval the SEC is violated. That suggests that the star will eventually evolve into a singularity, contrasting again with the likely outcome of the middle panel of Fig. \ref{min_pos_k_r_m}. However, a singularity can still be avoided if the collapsing object evaporates. If one chooses a very small initial radius, for instance, $r_0 =10^{-4}/a_b$, that can happen, as shown in the right panel of Fig. \ref{initial_conditions_sol2}. For the $(-)$-solution, there is no region in the space of parameters where all energy conditions are valid and the pressure and the energy density are positive simultaneously. This implies that the $(-)$-solution cannot describe a gravitational collapse when $a$ and $\phi$ are Taylor-expanded. Possibly, a fully numerical study might lead to a solution. For the $(\pm)$-solutions, the energy density is positive for all values of $t$ and the NEC is always satisfied.

\begin{figure}[ht]
\centering
    \includegraphics[scale=0.35]{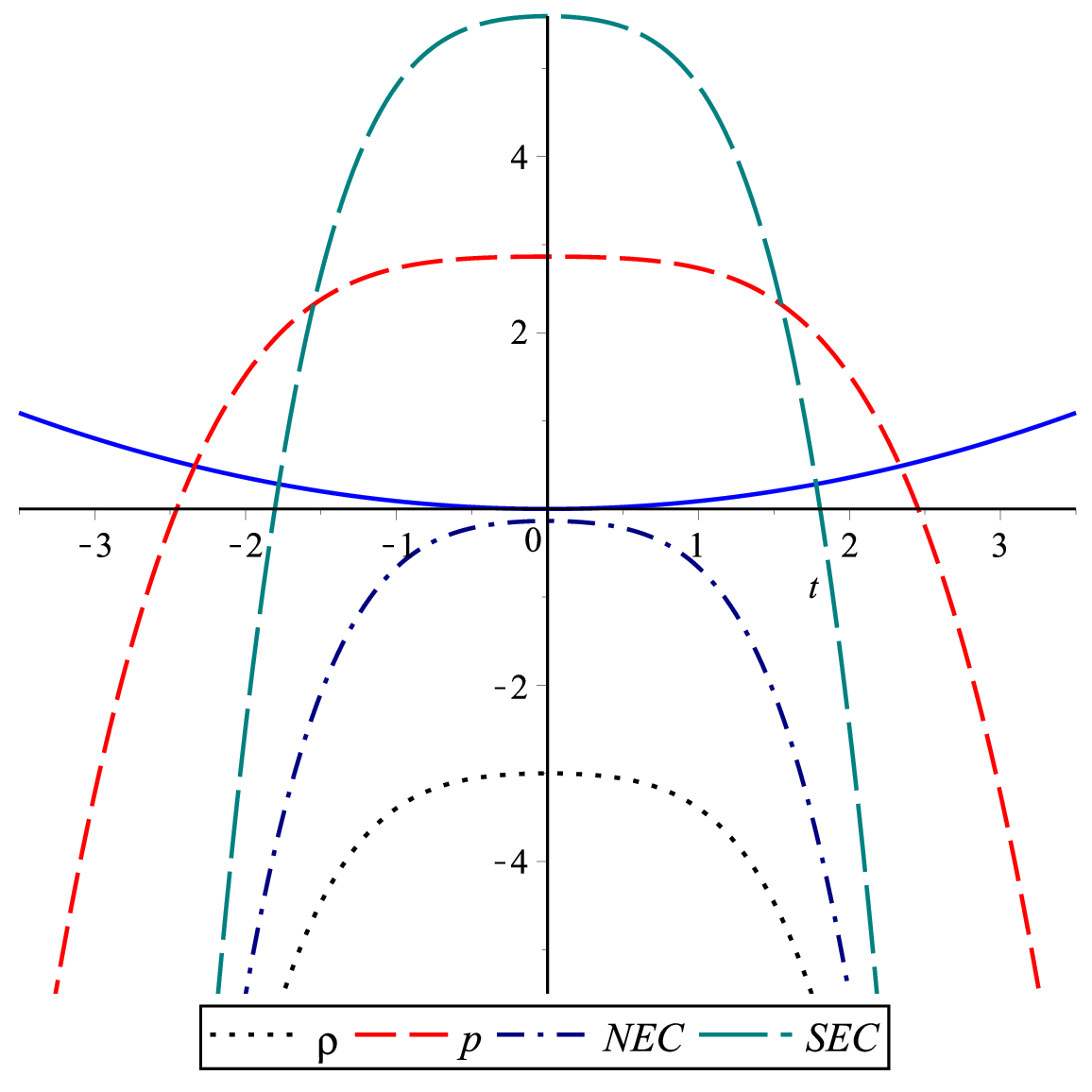}
    \includegraphics[scale=0.35]{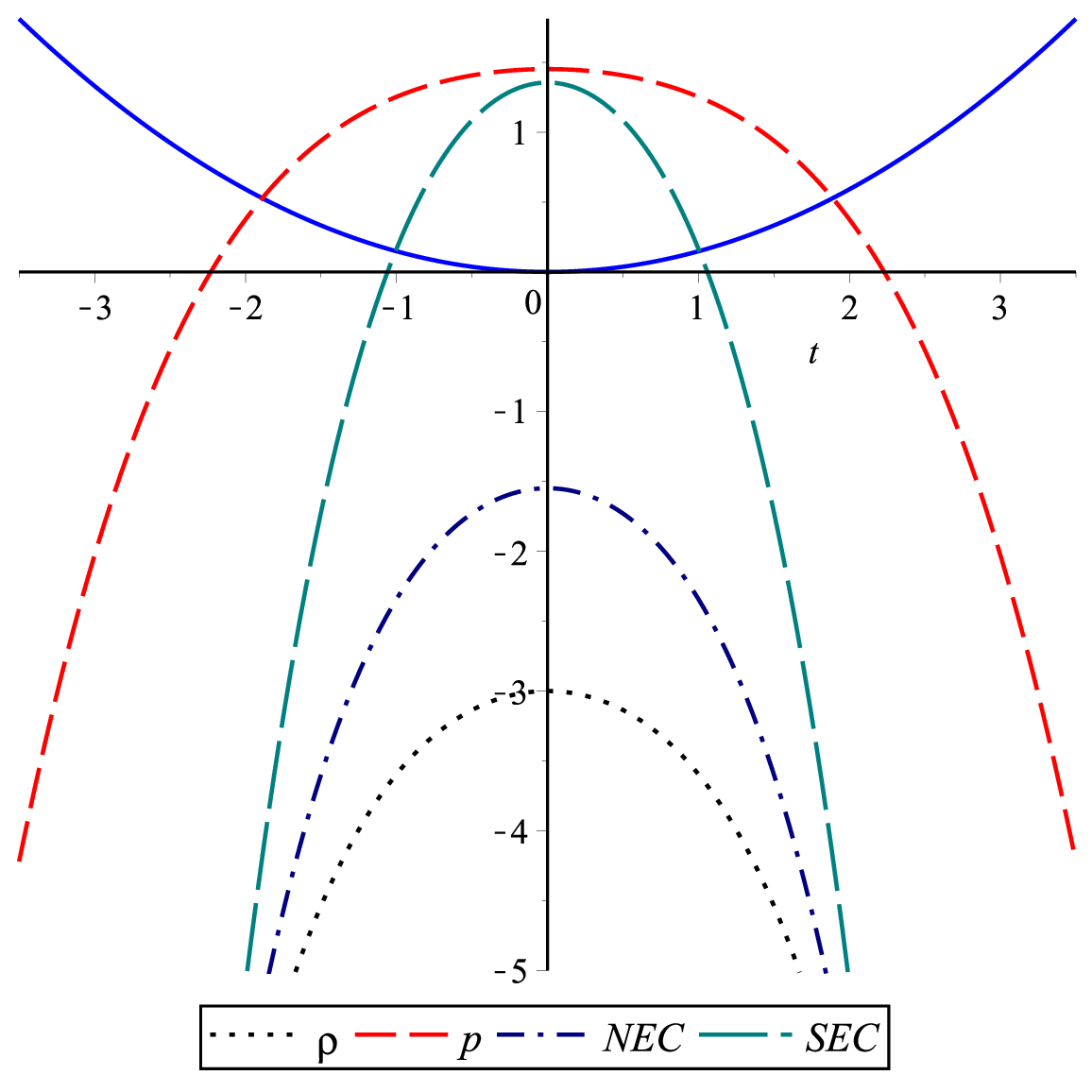}
    \caption{Behavior of the energy density, pressure, and energy conditions as a function of the time for the $(+)$-solution (on the left) and for the $(-)$-solution (on the right), with $k=-1$. In the interval where the solid curves are below the unit, we have the domains of validity of the Taylor expansion of $a$ and $\phi$. There is no time for which all the physical quantities are positive and the energy conditions are fulfilled.}
    \label{k_neg_non_minimal}
\end{figure}

For the case $k=-1$, the energy density is always negative and the NEC is never satisfied. Therefore, its physical meaning is not clear. For the sake of illustration, in Figure \ref{k_neg_non_minimal} we provide the solutions $\rho(t)$ and $p(t)$, and the behavior of the energy conditions. For both plots, one sees that the energy density and the NEC are negative and not fulfilled, respectively, while there are small intervals for which the pressure is positive and the SEC is satisfied. The solid blue line indicates the domain of validity of the power series expansion for these cases.

\section{Concluding remarks}
\label{conclusions}
We studied the viability of a gravitational collapse driven by a scalar field minimally and non-minimally coupled to gravity. The idea was to see the role different energy conditions play in the singularity theorems. The general conclusion is that if some hypothesis of the singularity theorems are violated, even for some intervals of time, the evolution of a gravitational collapse cannot be predicted a priori; all situations are possible. We have shown examples where there is evaporation, bounce, and the avoidance of singularities when the SEC is violated for both the minimal and non-minimal couplings.

It is worth noting that despite somewhat similar possibilities of singularity avoidance for both the minimal and non-minimal couplings, the contexts for so are very different. In the non-minimal case, only $k=1$ allows that, while it can happen for $k=\pm 1$ in the minimal case. The behavior of $\phi$ are totally different concerning the type of coupling. In the minimal case, it is dynamic, while in the non-minimal case, it is static ($\phi_0$). These different behaviors of $\phi$ depending on the coupling play an important role in the eventual violation of the energy conditions, in particular the SEC. 
Our analysis shows that the minimal coupling leads to a more ``general'' outcome, where all possibilities for the end result of a gravitational collapse happen. The non-minimal coupling is more restrictive. All the above makes it clear that both minimal and non-minimal couplings are viable possibilities for describing a gravitational collapse but can have different outcomes, which suggests only observations could tell which one is preferred. 

We finally note that although the internal spacetime is not accessible to external observers, who, thus, cannot access its aspects, part of this information is available at the hypersurface limiting it. The initially collapsing and then bounce behavior of the surface of a gravitational object is an indication to external observers that the SEC is violated in its interior (for an interval of time or indefinitely). Our analyses suggest that it could be interpreted as the existence of scalar fields there. 

\acknowledgments
The authors are thankful for the support of CNPq (E.B. grant N.\ 305217/2022-4), CAPES (A.G.C. grants N. 88887.501470/2020-00 and 88887.666979/2022-00), and FAPES (J.P.P. grant N. 04/2022).

\bibliography{ref}

\end{document}